\documentclass{article}
\usepackage{spconf,amsmath,graphicx}
\usepackage{amssymb}
\usepackage{enumitem}

\usepackage[table]{xcolor}
\usepackage{caption}
\usepackage{subcaption}
\usepackage{moresize}
\usepackage{multirow}
\usepackage{algorithm}
\usepackage{algpseudocode}
\usepackage{tikz}
\usepackage{url}

\usepackage{pgfplots}
\pgfplotsset{compat=1.8}
\pgfplotsset{colormap/hot2}

\definecolor{mycolor1}{rgb}{0.00000,0.44700,0.74100}%
\definecolor{mycolor2}{rgb}{0.85000,0.32500,0.09800}%
\definecolor{mycolor3}{rgb}{0.92900,0.69400,0.12500}%
\definecolor{mycolor4}{rgb}{0.46667,0.67451,0.18824}%
\definecolor{mycolor5}{rgb}{0.49412,0.18431,0.30588}%
\definecolor{mycolor6}{rgb}{0.30196,0.69020,0.93333}%

\usepgfplotslibrary{colorbrewer}

\title{SUBSPACE-BASED FEATURE FUSION FROM HYPERSPECTRAL AND MULTISPECTRAL IMAGES FOR LAND COVER CLASSIFICATION}
\name{Juan Ram\'irez$^{\dagger}$, H\'ector Vargas$^{\star}$, Jos\'e Ignacio Mart\'inez$^{\dagger}$, Henry Arguello$^{\ddag}$ \thanks{This work has received funding from the European Union’s Horizon 2020 research and innovation programme under the Marie Skłodowska-Curie grant agreement No 754382, GOT ENERGY TALENT. The content of this article does not reflect the official opinion of the European Union. Responsibility for the information and views expressed herein lies entirely with the authors.}
\thanks{\textcopyright 2021 IEEE. Personal use of this material is permitted. Permission from IEEE must be obtained for all other uses, in any current or future media, including reprinting/republishing this material for advertising or promotional purposes, creating new collective works, for resale or redistribution to servers or lists, or reuse of any copyrighted component of this work in other works.}}
\address{$^{\dagger}$Universidad Rey Juan Carlos, Móstoles, Comunidad de Madrid, Spain. \\ $^{\star}$Universidad Manuela Beltr\'an, Bogot\'a, Colombia. \\ $^{\ddag}$Universidad Industrial de Santander, Bucaramanga, Colombia.}
\begin{document}
\ninept
\maketitle

\begin{abstract}
\vspace{-1mm} In remote sensing, hyperspectral (HS) and multispectral (MS) image fusion have emerged as a synthesis tool to improve the data set resolution.  However, conventional image fusion methods typically degrade the performance of the land cover classification. In this paper, a feature fusion method from HS and MS images for pixel-based classification is proposed. More precisely, the proposed method first extracts spatial features from the MS image using morphological profiles. Then, the feature fusion model assumes that both the extracted morphological profiles and the HS image can be described as a feature matrix lying in different subspaces. An algorithm based on combining alternating optimization (AO) and the alternating direction method of multipliers (ADMM) is developed to solve efficiently the feature fusion problem. Finally, extensive simulations were run to evaluate the performance of the proposed feature fusion approach for two data sets. In general, the proposed approach exhibits a competitive performance compared to other feature extraction methods.
\end{abstract}
\vspace{-2mm}
\section{Introduction}
\label{sec:intro}
\vspace{-2mm}

Hyperspectral (HS) images are three-dimensional (3-D) data sets that contain the information of bi-dimensional (2-D) scenes across a large number of spectral bands. In this sense, the rich spectral information provided by HS images has been used in different remote sensing applications to identify materials on land covers \cite{ghamisi2017advances}. More precisely, spectral image classification methods have exploited the information embedded in the spectral signatures to label materials and objects in the scene of interest \cite{camps2013advances}. However, to reach a high signal-to-noise ratio (SNR) in the acquired measurements, HS imaging sensors frequently scan the scene using low-spatial resolution cameras. In consequence, to obtain high-resolution spectral images, various methods that fuse HS images with high-spatial-resolution multispectral (MS) images have been developed \cite{yokoya2017hyperspectral}.

Supervised pixel-based classifiers are frequently affected by the redundant information embedded in high-resolution spectral images. In particular, this harmful effect is referred to as the curse of the dimensionality that causes overfitting of the classifier bacause of a limited availability of training samples. To overcome this limitation, various dimensionality reduction (DR) methods have been proposed based on the assumption that the relevant information of spectral images lying in a lower-dimensional subspace. Principal component analysis (PCA), the orthogonal total variation component analysis (OTVCA) \cite{rasti2016hyperspectral}, and the sparse and smooth low-rank analysis (SSLRA) \cite{rasti2019hyperspectral} are just few examples that extract low-dimensional features from high-resolution spectral images. On the other hand, the fusion of multi-sensor remote sensing data can help to extract more information about the same land cover features, and enhance the classification accuracy. For example, a subspace-based multisensor fusion method has been reported in \cite{rasti2020remote} that characterizes multimodal data in different low-dimensional subspaces. 

In this paper, a feature fusion framework is developed to merge spatial-spectral information provided by HS and MS images for land cover classification. For this, the proposed framework first extracts spatial features from the MS image by applying a sequence of opening and closing by reconstruction operators with increasing-size structuring elements. Then, the fusion problem is modeled using a reduced-dimensional representation by linear projection into two different subspaces. By incorporating spatial information, the numerical procedure minimizes the total variational (TV) of the feature coefficients subject to an orthogonality constraint for the feature space. It is worth noting that, in contrast to the work reported in \cite{rasti2020remote}, the feature fusion formulation considers the spatial degradation affecting the HS data. Using spatial degradation operator can considerably improve the classification maps, resulting in improved detection of the borders of spatial structures. The numerical procedure is based on combining alternating optimization (AO) and the alternating direction method of multipliers (ADMM). Additionally, supervised pixel-based classification method is applied to the low-dimensional fused features to obtain the labeling maps. Extensive simulations were run to evaluate the performance of the classification framework for two data sets.

This paper is organized as follows. Section \ref{sec:model} describes the feature fusion problem from HS and MS images. The feature fusion algorithm based AO-ADMM is developed in Section \ref{sec:algorithm} and Section \ref{sec:results} shows the simulation results for two spectral image data sets. Some concluding remarks are summarized in Section \ref{sec:conclusions}.

\section{Problem Statement}
\label{sec:model}

First, it is necessary to introduce some notation. The transpose of a matrix is denoted by $\boldsymbol{X}^{\mathsf{T}}$. The $i$-th row of the matrix $\boldsymbol{X}$ is represented as $\boldsymbol{X}_{(i,:)}$, in the same way, for $j$-th column is denoted as $\boldsymbol{X}_{(:,j)}$. The expression $\boldsymbol{I}_{\!n}$ means the identity matrix of size $n \times n$. 

\textbf{\textit{Feature Fusion Model} :} Let $\boldsymbol{Y}_{\!\text{h}} \in \mathbb{R}^{m \times N_{\lambda}}$ be an observed HS image with $N_{\lambda}$ bands and $m = M_x \times M_y$ pixels, and $\boldsymbol{Y}_{\!\text{m}} \in \mathbb{R}^{n \times M_{\lambda}}$ an observed MS image with $M_{\lambda}$ bands and $n = N_x \times N_y$ pixels, with $m < n$ and $M_{\lambda} < N_{\lambda}$. Then, the observed HS and MS images can be characterized according to :
\vspace{-1mm}
\begin{equation}\label{equ:mult_fus}
            \boldsymbol{Y}_{\!\text{h}} = \boldsymbol{SK}\boldsymbol{Z} + \boldsymbol{H}_{\text{h}}, \quad
            \boldsymbol{Y}_{\!\text{m}}
              = \boldsymbol{Z}\boldsymbol{R}+ \boldsymbol{H}_{\text{m}},
\vspace{-1mm}
\end{equation}
where $\boldsymbol{Z} \in \mathbb{R}^{n \times N_{\lambda}}$ is the target high-resolution image, $\boldsymbol{S} \in \mathbb{R}^{m \times n}$ is a down-sampling matrix with factor $d$ (with $n = d^2m$), $\boldsymbol{K} \in \mathbb{R}^{n \times n}$ is a periodic spatial blurring operator, $\boldsymbol{R} \in \mathbb{R}^{N_{\lambda} \times M_{\lambda}}$ is the spectral response of the MS sensor. $\boldsymbol{H}_{\text{h}} \in \mathbb{R}^{m \times N_{\lambda}}, \boldsymbol{H}_{\text{m}} \in \mathbb{R}^{n \times M_{\lambda}}$ are additive terms that include both modeling errors and sensor noises. Note that, the spatial blur $\boldsymbol{K}$ and the spectral response $\boldsymbol{R}$ can be estimated from observed images $\boldsymbol{Y}_{\!\text{h}}$ and $\boldsymbol{Y}_{\!\text{m}}$ \cite[sec.~IV]{simoes2014convex}. 

In essence, the image fusion problem consists of estimating the high-resolution image $\boldsymbol{Z}$, given the observed data $\boldsymbol{Y}_{\!\text{h}}$ and $\boldsymbol{Y}_{\!\text{m}} $ using the model (\ref{equ:mult_fus}) \cite{wei2015fast,wei2016multiband}. Since the bands of the HS image are generally dependent, those methods assume that  $\boldsymbol{Z}$ usually lives in a subspace whose dimension $N_e$ is much smaller than the number of bands $N_{\lambda}$. In contrast to the image fusion techniques, the aim of this work is to estimate to estimate high-resolution features from observed images $\boldsymbol{Y}_{\!\text{h}}$ and $\boldsymbol{Y}_{\!\text{m}}$. By considering first $\{\psi_i, \phi_i\}_{i = 1,\dots,p}$, a set of opening and closing by reconstruction operators, respectively, where $p$ refers to the size of the filter. The mathematical morphology of the $i$-th multispectral band, i.e, $\boldsymbol{y}_{\text{m},i} = \boldsymbol{Y}_{\!\text{m}(:,i)}$, using the latter set of operators is defined as:
\vspace{-1mm}
\begin{equation*}
    \boldsymbol{Y}_{\!i} = \left[ \psi_p(\boldsymbol{y}_{\text{m},i}) \cdots \psi_1(\boldsymbol{y}_{\text{m},i}) \mbox{ } \boldsymbol{y}_{\text{m},i} \mbox{ } \phi_1(\boldsymbol{y}_{\text{m},i}) \cdots \phi_p(\boldsymbol{y}_{\text{m},i}) \right],
\vspace{-1mm}
\end{equation*}
where $\boldsymbol{Y}_{\!i} \in \mathbb{R}^{n \times 2p + 1}$ is the matrix contains the vectorized morphological features \cite{benediktsson2005classification}. 
Then, if $\boldsymbol{Y}_{\!\text{mp}} = [\boldsymbol{Y}_{\!1} \mbox{ } \cdots \mbox{ } \boldsymbol{Y}_{\!M_{\lambda}} ]$, the features from $\boldsymbol{Y}_{\!\text{h}}$ and $\boldsymbol{Y}_{\!\text{mp}}$ can be modeled in lower dimensional space as
\vspace{-1mm}
\begin{equation}\label{equ:mult_fus_feat}
    \boldsymbol{Y}_{\!\text{h}}  = \boldsymbol{SK}\boldsymbol{C}\boldsymbol{Q}+ \boldsymbol{H}_{\text{h}}, 
    \quad
    \boldsymbol{Y}_{\!\text{mp}} = \boldsymbol{C}\boldsymbol{Q}_{\text{mp}} + \boldsymbol{H}_{\text{mp}},
    \vspace{-1mm}
\end{equation}
where $\boldsymbol{C} \in \mathbb{R}^{n \times N_e}$ are the joint coefficients, $\boldsymbol{Q} \in  \mathbb{R}^{N_e \times N_{\lambda}}$, $\boldsymbol{Q}_{\text{mp}} \in  \mathbb{R}^{N_e \times M_{\lambda}}$ are subspace matrices, and $\boldsymbol{H}_{\text{h}}$, $\boldsymbol{H}_{\text{mp}}$ represent the noise and model error. The matrices $\boldsymbol{Q}$, $\boldsymbol{Q}_{\text{mp}}$ are assumed orthonormal matrices such that $\boldsymbol{Q}\boldsymbol{Q}^\mathsf{T} = \boldsymbol{Q}_{\text{mp}}\boldsymbol{Q}_{\text{mp}}^\mathsf{T} = \boldsymbol{I}_{\!N_e}$.



\textbf{\textit{Constrained Optimization} :} Under the assumption that noise matrices in (\ref{equ:mult_fus_feat}) obey to a Gaussian distribution, the estimation of the matrices $\boldsymbol{C}$, $\boldsymbol{Q}$ from matrices $\boldsymbol{Y}_{\!\text{h}}$, $\boldsymbol{Y}_{\!\text{mp}}$ can be obtained by solving the following constrained optimization problem
\vspace{-1mm}
\begin{equation}\label{equ:obj_fnc_tv}
        \min_{\substack{\boldsymbol{C}, \boldsymbol{Q} }}  \quad J\left(\boldsymbol{C}, \boldsymbol{Q} \right),  \quad \text{s.t.}  \quad \boldsymbol{Q} \boldsymbol{Q}^{\mathsf{T}} = \boldsymbol{I}_{\!N_e}
        \vspace{-2mm}
\end{equation}
where
\vspace{-2mm}
\begin{equation*}
    \begin{split}
        J(\boldsymbol{C}, \boldsymbol{Q} ) & = \frac{1}{2} \| \boldsymbol{SK}\boldsymbol{C}\boldsymbol{Q}  -\boldsymbol{Y}_{\!\text{h}}\|_{\mathsf{F}}^2 + \frac{\lambda}{2} \| \boldsymbol{C}\boldsymbol{Q}_{\text{mp}} - \boldsymbol{Y}_{\!\text{mp}} \|_{\mathsf{F}}^2 + \dots\\
        & \qquad \lambda_{\text{TV}} \sum_{i=1}^{N_e} \left\| \begin{bmatrix} \boldsymbol{D}_{\text{h}}\boldsymbol{C}_{(:,i)}  \mbox{ } \boldsymbol{D}_{\text{v}}\boldsymbol{C}_{(:,i)} \end{bmatrix}  \right\|_{2,1}
    \end{split},
    \vspace{-2mm}
\end{equation*}
$\| \boldsymbol{X} \|_{2,1} = \sum_{i=1}^n \| \boldsymbol{X}_{(i,:)} \|_2$ is the mixed $\ell_{2,1}$ norm, and $\| \cdot \|_{\mathsf{F}}$ is
the Frobenius norm. The matrices $\boldsymbol{D}_{\text{h}}, \boldsymbol{D}_{\text{v}} \in \mathbb{R}^{n \times n}$ are operators to calculate the first order vertical and horizontal differences, respectively, for a vectorized image. Note that, the formulation in (\ref{equ:obj_fnc_tv}) exploits TV penalty and an orthogonality constraint. The TV penalty ensures some spatial smoothness and preserves edges, which are boundaries of objects that are used to improve classification procedures.

\begin{algorithm}
	\footnotesize
    \caption{Feature fusion based on AO-ADMM.}
	\label{alg:ao_admm}
	\hspace*{\algorithmicindent} \textbf{input:} $\boldsymbol{Y}_{\!\text{h}}$, $\boldsymbol{Y}_{\!\text{mp}}$, $\boldsymbol{K}$, $N_e$, $\lambda$ and $\lambda_{\text{TV}}$. 		
	\begin{algorithmic}[1]
		\Statex{// \texttt{initialize } $\boldsymbol{C}_2^{(0)}$, $\boldsymbol{C}_3^{(0)}$ \texttt{ with zeros.} }
		\State{$\boldsymbol{Q}^{(0)} \gets  (\boldsymbol{V}_{\!(:,1:N_e)})^{\mathsf{T}}$  {\hspace{2mm}// \texttt{where} $[\boldsymbol{U}, \boldsymbol{S}, \boldsymbol{V}] = \text{svd}(\boldsymbol{Y}_{\!\text{h}}^{\mathsf{T}}\boldsymbol{Y}_{\!\text{h}})$} }
		\State{$\boldsymbol{Q}_{\text{mp}} \gets  (\boldsymbol{V}_{\!\text{mp}(:,1:N_e)})^{\mathsf{T}}$ {//   \texttt{where} $[\boldsymbol{U}_{\text{mp}}, \boldsymbol{S}_{\text{mp}}, \boldsymbol{V}_{\!\text{mp}}] = \text{svd}(\boldsymbol{Y}_{\!\text{mp}}^{\mathsf{T}}\boldsymbol{Y}_{\!\text{mp}})$} }
		\For{$t= 1,2,\dots$ \textbf{to} \textit{stopping rule} }
			\Statex{// \texttt{Optimize $\boldsymbol{C}$ using ADMM (see Alg.(2)). }} 				
			\State{$ [ \boldsymbol{C}^{(t)},\boldsymbol{C}_2^{(t)},\boldsymbol{C}_3^{(t)}] =  \text{ADMM}(\boldsymbol{Y}_{\!\text{h}}, \boldsymbol{Y}_{\!\text{mp}}, \boldsymbol{K}, \boldsymbol{Q}_{\text{mp}}, \lambda, \lambda_{\text{TV}}, \dots$}
			\Statex{$\qquad \qquad \qquad \qquad \qquad \qquad \quad \quad \boldsymbol{Q}^{(t-1)},\boldsymbol{C}_2^{(t-1)}, \boldsymbol{C}_3^{(t-1)})$}
			\State{$\boldsymbol{Q}^{(t)} \gets \boldsymbol{U} \boldsymbol{V}^{\mathsf{T}}$ {\hspace{2mm}//   \texttt{where} $[\boldsymbol{U}, \boldsymbol{S}, \boldsymbol{V}] = \text{svd}((\boldsymbol{SKC}^{(t)})^{\mathsf{T}}\boldsymbol{Y}_{\!\text{h}})$} }
			\EndFor
		\end{algorithmic}
		\hspace*{\algorithmicindent} set $\hat{\boldsymbol{C}} = \boldsymbol{C}^{(t)}$ and \textbf{output:}  $\hat{\boldsymbol{C}}$  
\end{algorithm}	 

\section{Alternating Optimization Scheme}
\label{sec:algorithm}

The problem in (\ref{equ:obj_fnc_tv}) is solved one matrix at a time, while the other is assumed to be fixed. This procedure is summarized in Algorithm \ref{alg:ao_admm}, where the AO estimator is adopted to solve efficiently $\boldsymbol{C}$ and $\boldsymbol{Q}$ iteratively. To overcome the closed-form expression problem in (\ref{equ:obj_fnc_tv}), the ADMM is embedded in each iteration of the AO algorithm. The convergence analysis and stopping rule for Algorithm \ref{alg:ao_admm} are similar to the algorithm presented in \cite{vargas2019low}. 

\begin{algorithm}
    \footnotesize
	\caption{ ADMM sub-iterations to estimate $\boldsymbol{C}$}	
	\label{alg:admm_c}	
	\begin{algorithmic}[1]	
			\Function{ADMM}{$\boldsymbol{Y}_{\!\text{h}}, \boldsymbol{Y}_{\!\text{mp}}, \boldsymbol{K}, \boldsymbol{Q}_{\!\text{mp}}, \lambda, \lambda_{\text{TV}},\boldsymbol{Q}, \boldsymbol{U}, \boldsymbol{G}$}	
			\State{$\boldsymbol{U}^{(0)}= \boldsymbol{U},\boldsymbol{G}^{(0)}=\boldsymbol{G}, \rho>0$}
			\For{$t= 1,2,\dots$ \textbf{to} \textit{stopping rule} }
			\State{$\boldsymbol{C}^{(t)} \in \text{ $\underset{\boldsymbol{C}}{\operatorname{argmin}}$ } \mathcal{L}(\boldsymbol{C}, \boldsymbol{U}^{(t-1)},\boldsymbol{G}^{(t-1)})$}
			\hspace{2mm}// \texttt{Eq.(\ref{equ:uptg_c})}
			\State{$\boldsymbol{U}^{(t)} \in \text{ $\underset{\boldsymbol{U}}{\operatorname{argmin}}$ } \mathcal{L}(\boldsymbol{C}^{(t)}, \boldsymbol{U}, \boldsymbol{G}^{(t-1)})$}
			\hspace{2mm}// \texttt{Eq.(\ref{equ:uptg_u})}
			\State{$\boldsymbol{G}^{(t)} = \boldsymbol{G}^{(t-1)} + \boldsymbol{AC}^{(t)} + \boldsymbol{BU}^{(t)} $}
			\EndFor
			\State{set $\boldsymbol{C} = \boldsymbol{C}^{(t)}$, $\boldsymbol{U} = \boldsymbol{U}^{(t)}$, $\boldsymbol{G} = \boldsymbol{G}^{(t)}$ and \Return {$\boldsymbol{C}, \boldsymbol{U}, \boldsymbol{G}$}}
			\EndFunction					
	\end{algorithmic}
\end{algorithm}

\textbf{\textit{Fix}} $\boldsymbol{Q}$ \textbf{\textit{and Updating}} $\boldsymbol{C}$: Given a fixed $\boldsymbol{Q}$, the minimization problem in (\ref{equ:obj_fnc_tv}) can be solved by introducing an auxiliary variable $\boldsymbol{U}$, splitting the objective, and using the ADMM method. The optimization problem in (\ref{equ:obj_fnc_tv}) with respect to $\boldsymbol{C}$ can be written as:
\vspace{-1mm}
\begin{equation}\label{equ:c_prob}
        \min_{\substack{\boldsymbol{C}, \boldsymbol{U} }}  \quad f(\boldsymbol{C}) + g(\boldsymbol{U}),  \quad \text{s.t.}  \quad \boldsymbol{AC} + \boldsymbol{BU} = \boldsymbol{0}
        \vspace{-2mm}
\end{equation}
where 
\vspace{-2mm}
\begin{equation*}
\begin{split}
    g(\boldsymbol{U}) & = \lambda_{\text{TV}} \sum_{i=1}^{N_e} \left\| \begin{bmatrix} \boldsymbol{U}_{\!1(:,i)} \mbox{ } \boldsymbol{U}_{\!2(:,i)} \end{bmatrix}  \right\|_{2,1} \\
    f(\boldsymbol{C}) & = \frac{1}{2} \| \boldsymbol{SK}\boldsymbol{C}\boldsymbol{Q} - \boldsymbol{Y}_{\!\text{h}}\|_{\mathsf{F}}^2 + \frac{\lambda}{2} \| \boldsymbol{C}\boldsymbol{Q}_{\text{mp}}  -\boldsymbol{Y}_{\!\text{mp}}  \|_{\mathsf{F}}^2 \\
    \boldsymbol{A} & = \begin{bmatrix} \boldsymbol{D}_{\mathsf{h}}\\ \boldsymbol{D}_{\mathsf{v}} \end{bmatrix}, \quad \boldsymbol{B}  = \begin{bmatrix} -\boldsymbol{I}_{\!n} & \boldsymbol{0} \\ \boldsymbol{0} & -\boldsymbol{I}_{\!n}  \end{bmatrix}, \quad \boldsymbol{U} = \begin{bmatrix} \boldsymbol{U}_{\!1} \\ \boldsymbol{U}_{\!2} \end{bmatrix}
\end{split},
\vspace{-1mm}
\end{equation*}
and $\boldsymbol{U}_{\!1},\boldsymbol{U}_{\!2} \in \mathbb{R}^{n \times N_e}$. The iterative procedure to solve the formulation in (\ref{equ:c_prob}) is shown in Algorithm \ref{alg:admm_c}. The  augmented Lagrangian function is defined as
\vspace{-1mm}
\begin{equation}\label{equ:lag_admm}
    \mathcal{L}(\boldsymbol{C}, \boldsymbol{U}, \boldsymbol{G}) = f(\boldsymbol{C}) + g(\boldsymbol{U}) + \frac{\rho}{2}\| \boldsymbol{AC} + \boldsymbol{BU} + \boldsymbol{G}\|_{\mathsf{F}}^2,
    \vspace{-1mm}
\end{equation}
where $\rho > 0$ is the Lagrange penalty parameter \cite{boyd2011distributed}, and $\boldsymbol{G} = [\boldsymbol{G}_1; \boldsymbol{G}_2]$ with $\boldsymbol{G}_1,\boldsymbol{G}_2 \in \mathbb{R}^{n \times N_e}$ denotes the scaled Lagrange multipliers related to the constraint $\boldsymbol{AC} + \boldsymbol{BU} = \boldsymbol{0}$. Then, $\mathcal{L}(\boldsymbol{C}, \boldsymbol{U}, \boldsymbol{G})$ is minimized with respect to $\boldsymbol{C}$ and $\boldsymbol{U}$, and update $\boldsymbol{G}$ as in Algorithm \ref{alg:admm_c}. Forcing the derivative of (\ref{equ:lag_admm}) with respect to $\boldsymbol{C}$ to be zero leads the following linear system:
\vspace{-1mm}
\begin{equation}\label{equ:uptg_c}
    \boldsymbol{C} \gets \left( (\boldsymbol{SK})^{\mathsf{T}}\boldsymbol{SK}  + \lambda\boldsymbol{I}_{\!n} + \rho(\boldsymbol{D}_{\mathsf{h}}^{\mathsf{T}}\boldsymbol{D}_{\mathsf{h}} + \boldsymbol{D}_{\mathsf{v}}^{\mathsf{T}}\boldsymbol{D}_{\mathsf{v}}) \right)^{-1} \boldsymbol{E}_1,
    \vspace{-1mm}
\end{equation}
where $\boldsymbol{E}_1 = (\boldsymbol{SK})^{\mathsf{T}}\boldsymbol{Y}_{\!\text{h}}\boldsymbol{Q}^{\mathsf{T}} + \lambda (\boldsymbol{Y}_{\!\text{mp}}\boldsymbol{Q}_{\text{mp}}^{\mathsf{T}}) + \rho(\boldsymbol{D}_{\mathsf{h}}^{\mathsf{T}}(\boldsymbol{U}_{\!1} - \boldsymbol{G}_1) + \boldsymbol{D}_{\mathsf{v}}^{\mathsf{T}}(\boldsymbol{U}_{\!2} - \boldsymbol{G}_2))$. By assuming first that, $\boldsymbol{K}$ is the matrix representation of the cyclic convolution operator, i.e., $\boldsymbol{K}$ is a block circulant matrix with circulant blocks. Furthermore, if $\boldsymbol{S}$ is a down-sampling operator, while its conjugate transpose $\boldsymbol{S}^{\mathsf{T}}$ interpolates the decimated image with zeros. The inversion of (\ref{equ:uptg_c}) can be solved analytically by following the Theorem 1 presented in \cite{zhao2016fast}. The sub-problem $\boldsymbol{U}$ is decouple into two variables in step 5 of Algorithm \ref{alg:admm_c}. The solution of $\boldsymbol{U}_{\!1}$ and $\boldsymbol{U}_{\!2}$ are given by 
\vspace{-1mm}
\begin{equation}\label{equ:uptg_u}
   \left[\boldsymbol{U}_{\!1(:,i)} \mbox{ }\boldsymbol{U}_{\!2(:,i)}\right] \gets \text{vect-soft}(\left[\boldsymbol{E}_{2(:,i)}\mbox{ } \boldsymbol{E}_{3(:,i)}\right], \lambda/\rho),
   \vspace{-1mm}
\end{equation}
where $\boldsymbol{E}_2 = \boldsymbol{D}_{\text{h}}\boldsymbol{C}+ \boldsymbol{U}_{\!1}$, $\boldsymbol{E}_3 = \boldsymbol{D}_{\text{v}}\boldsymbol{C} + \boldsymbol{U}_{\!2}$, and $\text{vect-soft}(\cdot, \tau)$ denotes
the vect-soft threshold function $\boldsymbol{x} \gets \boldsymbol{x}( \text{max}(\|\boldsymbol{x}\|_2-\tau,0)/\text{max}(\|\boldsymbol{x}\|_2-\tau,0) + \tau )$, with $\tau$ being a threshold \cite{simoes2014convex,zhao2016fast}. For the termination of Algorithm \ref{alg:admm_c}, the stopping criterion described in \cite[sec.~3.3.1]{boyd2011distributed} is adopted. Then, the Algorithm \ref{alg:admm_c} ends when $r_{\text{res}}$ and $s_{\text{res}}$ are smaller than some threshold $\epsilon_{\text{admm}}$, where 
 \vspace{-1mm}
\begin{equation*}
    \begin{split}
       r_{\text{res}} & = \|  \boldsymbol{AC}^{(t)} + \boldsymbol{BU}^{(t)} \|_{\mathsf{F}}/\text{max}(\| \boldsymbol{AC}^{(t)}\|_{\mathsf{F}}, \| \boldsymbol{BU}^{(t)}\|_{\mathsf{F}}) \\
	   s_{\text{res}} & = \|\rho\boldsymbol{A}^{\mathsf{T}}\boldsymbol{B}(\boldsymbol{U}^{(t)} - \boldsymbol{U}^{(t-1)}) \|_{\mathsf{F}}/ \| \boldsymbol{A}^{\mathsf{T}}\boldsymbol{G}^{(t)} \|_{\mathsf{F}}
    \end{split},
     \vspace{-1mm}
\end{equation*}
are the relative primal residual and dual residual, respectively. 

\textbf{\textit{Fix}} $\boldsymbol{C}$ \textbf{\textit{and Updating}} $\boldsymbol{Q}$: Given a fixed $\boldsymbol{C}$, the minimization problem in (\ref{equ:obj_fnc_tv}) with respect to $\boldsymbol{Q}$ is reduced to a orthogonal Procrustes problem given by
\vspace{-1mm}
\begin{equation}\label{equ:q_prob}
    \min_{\boldsymbol{Q}} \quad \frac{1}{2} \| \boldsymbol{SKC}\boldsymbol{Q} -\boldsymbol{Y}_{\text{h}} \|_{\mathsf{F}}^2,  \quad \text{s.t.}  \quad \boldsymbol{Q} \boldsymbol{Q}^{\mathsf{T}} = \boldsymbol{I}_{\!N_e},
    \vspace{-1mm}
\end{equation}
whose solution is given by $\boldsymbol{Q} \gets \boldsymbol{U}\boldsymbol{V}^{\mathsf{T}}$, where the matrices $\boldsymbol{U}$, $\boldsymbol{V}$ are given by the singular value decomposition of $(\boldsymbol{SKC})^{\mathsf{T}}\boldsymbol{Y}$. 

\vspace{-2mm}
\section{Simulation results}
\label{sec:results}
\vspace{-2mm}
This section presents the numerical results of the proposed method for land cover classification and different data sets. The proposed scheme was implemented in MATLAB \footnote{The source code for the proposed subspace-based feature fusion approach can be downloaded from this link:
\url{https://github.com/JuanMarcosRamirez/featurefusion_igarss2021}}. The computer simulations were performed using a desktop architecture with an Intel Core i7 CPU, 3.00 GHz, 64-GB RAM, and Ubuntu 18.04 operating system. 

\textbf{\textit{Algorithms Parameters}:}
For the morphology profiles, the radius of disk-shaped for opening and closing operators were selected as $[10\mbox{ }20\mbox{ }50\mbox{ }100\mbox{ }200]$. In Algorithm \ref{alg:ao_admm}, the stooping threshold is set to $\epsilon_{\textrm{ao}} < 10^{-4}$. In Algorithm \ref{alg:admm_c}, the stooping threshold and the Lagrange penalty parameter are set to $\epsilon_{\textrm{admm}} < 10^{-4}$ and $\rho = 1$, respectively. The parameter $\rho$ is adapted by balancing the primal and dual residuals \cite[sec.~3.4.1]{boyd2011distributed}. 

\begin{figure}[h]
\begin{center}
    \begin{tabular}{c c}
    \hspace{-5pt}
    \begin{minipage}{.35\linewidth}\vspace{-11pt}
    \includegraphics[width=\linewidth]{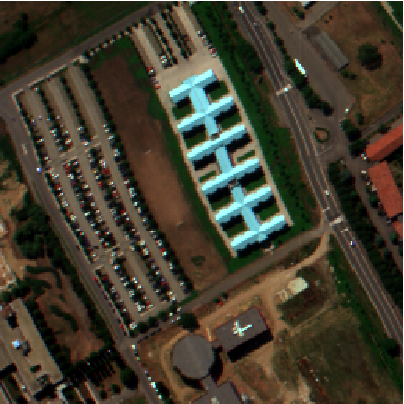}
    \end{minipage}
    &
    \begin{minipage}{.49\linewidth}
    \resizebox{1.10\textwidth}{!}{
\begin{tikzpicture}
    \begin{axis}[
    colorbar sampled,
    colorbar style={samples=16, width=0.1\textwidth,at={(rel axis cs: 2.00,1)}},
    3d box,
    xlabel = \large \textrm{$\lambda_{\mathrm{TV}}$},
    xmin = 0.0001, 
    xmax = 1, 
    xmode = log,
    ylabel = \large \textrm{$N_e$},
    ymin = 4, 
    zmin = 0.88, 
    zmax = 1, 
    ymax = 24,
    ticklabel style = {font=\large},
    view = {0}{90}]
    \addplot3[surf,mesh/ordering=y varies,
    shader=interp,  mesh/cols=10,contour filled={number=16}]table{FigData/OA_surface.dat};
    \addplot3 [densely dotted, smooth, line width = 1.00, mesh/cols=17, mesh/rows=10, contour gnuplot={number=16,labels={false},draw color=black}]table{FigData/OA_surface.dat};
    \end{axis}
\end{tikzpicture}}
    \end{minipage}
    \end{tabular}\vspace{-15pt}
\end{center}
    \caption{\footnotesize Pavia University data set, (left) the RGB composite of the original data set, and (right) the influence of $N_e$ and $\lambda_{\mathrm{TV}}$ on the overall accuracy. \vspace{-10pt}}
    \label{fig:real_data}
\end{figure}
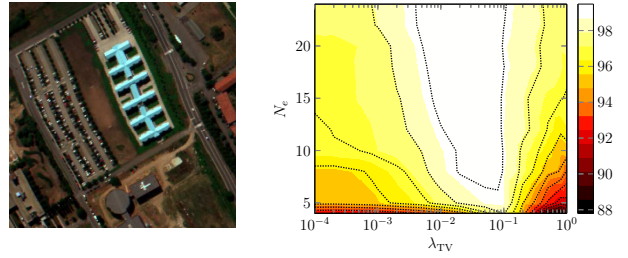

\textbf{\textit{Pavia University}:} was captured by the Reflective Optics Imaging Spectrometer (ROSIS) over Pavia, Italy. This image comprises 256 $\times$ 256 pixels with a spatial resolution of 1.3 m per pixel and 103 spectral channels in the wavelength range from  0.43 to 0.86 $\mu$m \cite{SpectralImageDatabase}. The RGB composite of the input image is shown in Fig. \ref{fig:real_data}(left). The HS image is obtained by downsampling every spectral channel of the original data set with a 4:1 spatial decimation factor. Hence, the HS image has dimensions of 64 $\times$ 64 pixels and 103 spectral bands. To emulate the imperfections during the acquisition process, we set the SNR of the HS image to 30 dB. On the other hand, the  MS image is obtained by projecting the original image along the spectral axis using the IKONOS spectral responses. The MS image has dimensions of $256 \times 256 \times 5$ and the SNR is fixed to $40$ dB. To evaluate the performance of the proposed feature fusion method for spectral image classification, we use a ground truth map with six different classes that identify distinct materials in an urban cover.


Note that the performance of the proposed feature fusion method depends on two parameters that need to be set, i.e. the number of fused features $N_{e}$ and the TV regularization parameter $\lambda _{\mathrm{TV}}$. In this regard, we realize an experiment on the Pavia University data set to evaluate the classification performance of the proposed method using different parameter settings. Fig. \ref{fig:real_data}(right) illustrates the overall accuracy induced by the proposed technique for different numbers of fused features and for different values of $\lambda _{\mathrm{TV}}$. More specifically, each point in the surface is obtained by averaging ten realizations of the respective experiment, and at each trial, we use a different noise component and a different training set is randomly selected. In this case, the number of training samples was fixed to 50 pixels per class and the classifier is a support vector machine (SVM) with a polynomial kernel of order three. As can be seen, the proposed approach exhibits a remarkable performance when $N_e > 5$. It can be also observed the influence of $\lambda _{\mathrm{TV}}$ in the evaluation interval.



\begin{figure}[h!]
\begin{center}
\begin{tabular}{c c c}
\hspace{-10pt}
\includegraphics[width=0.32\linewidth]{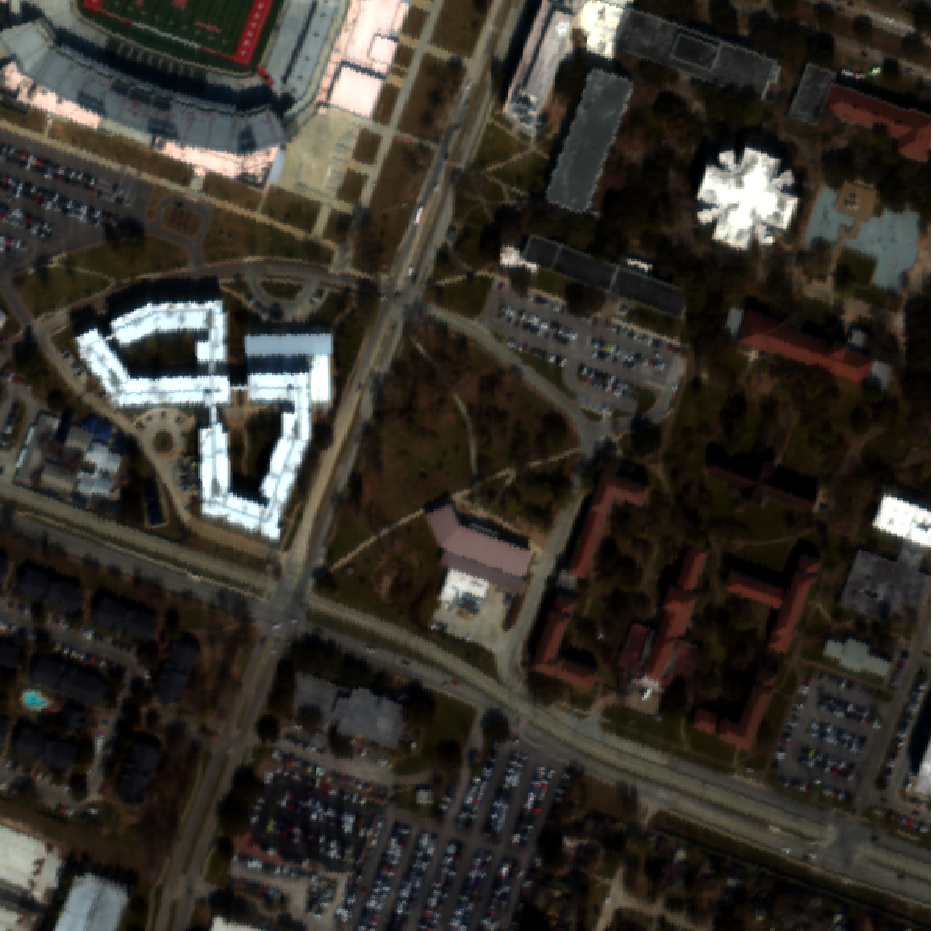}
&
\hspace{-10pt}
\includegraphics[width=0.32\linewidth]{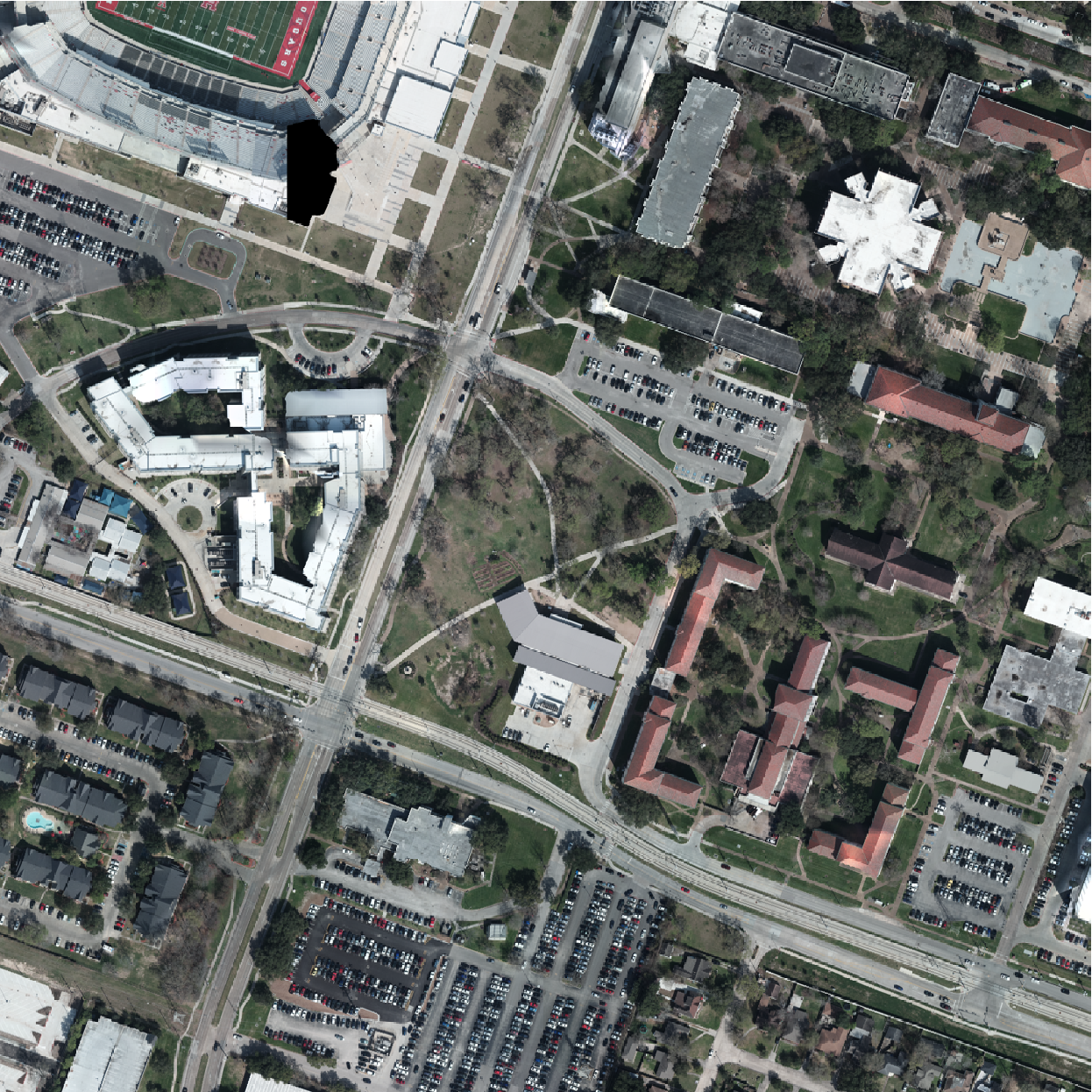}
&
\hspace{-10pt}
\includegraphics[width=0.32\linewidth]{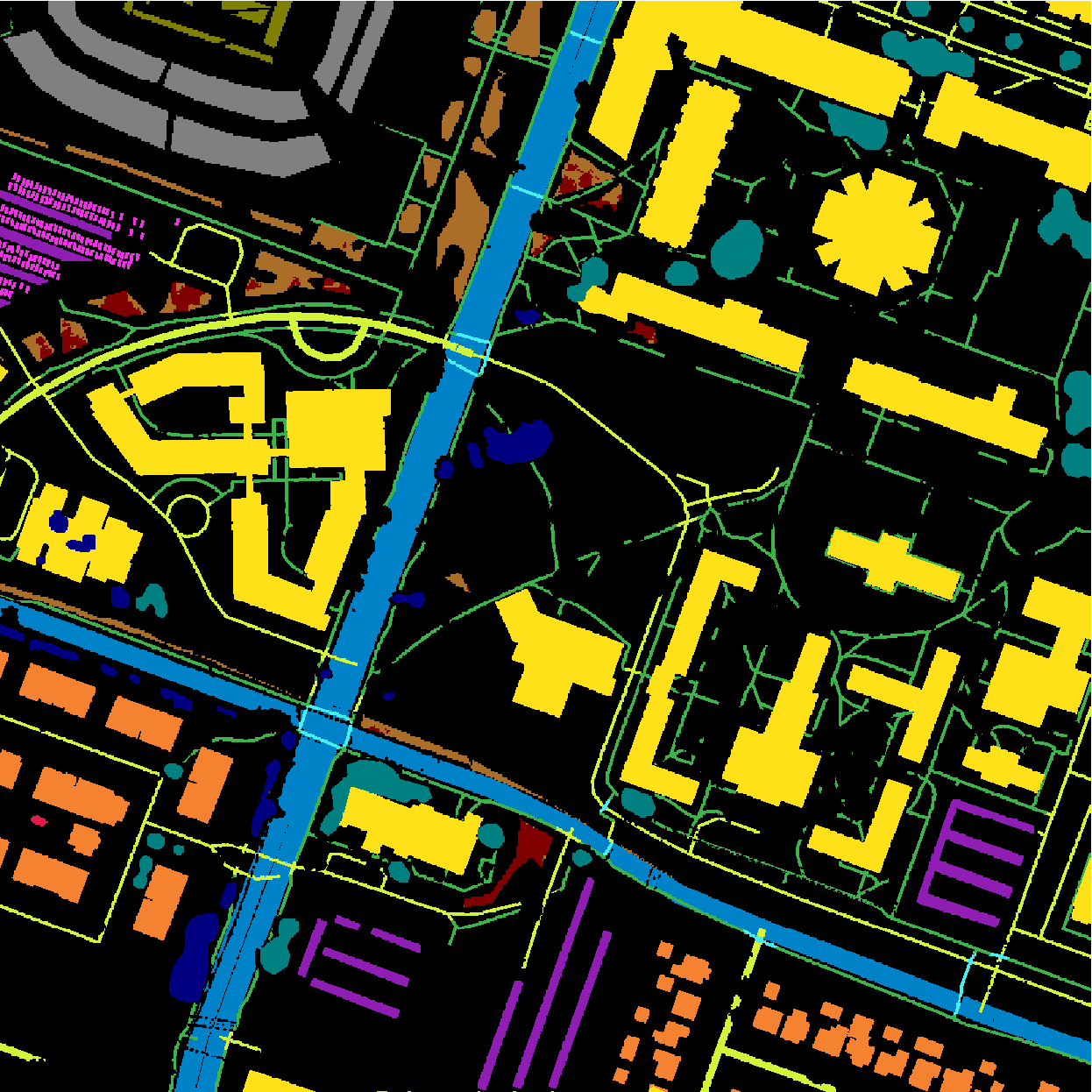}
\\
\hspace{-10pt} \scriptsize (a) HS image
&
\hspace{-10pt} \scriptsize (b) RGB image
&
\hspace{-10pt} \scriptsize (c) Ground truth
\\
\hspace{-10pt}
\includegraphics[width=0.32\linewidth]{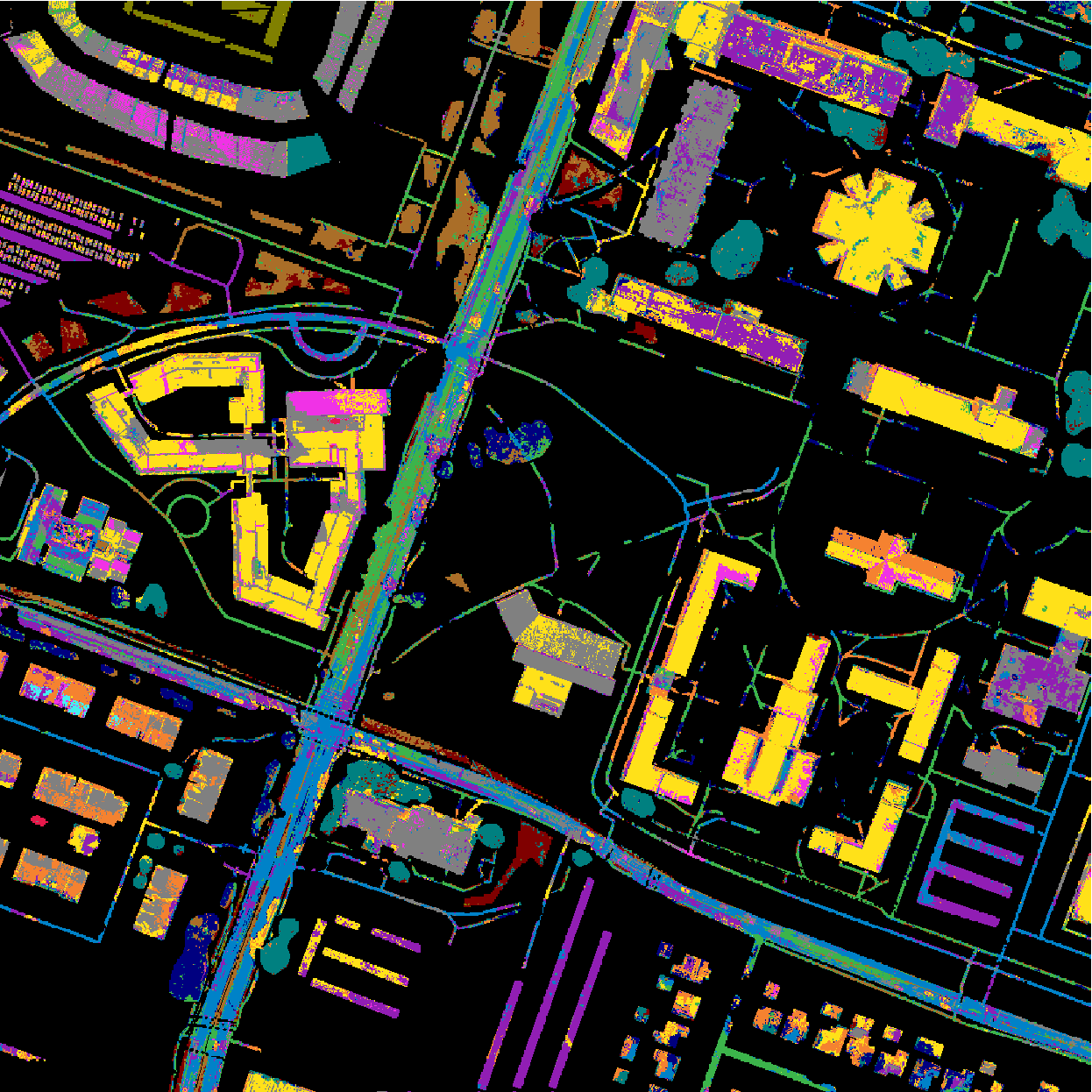}
&
\hspace{-10pt}
\includegraphics[width=0.32\linewidth]{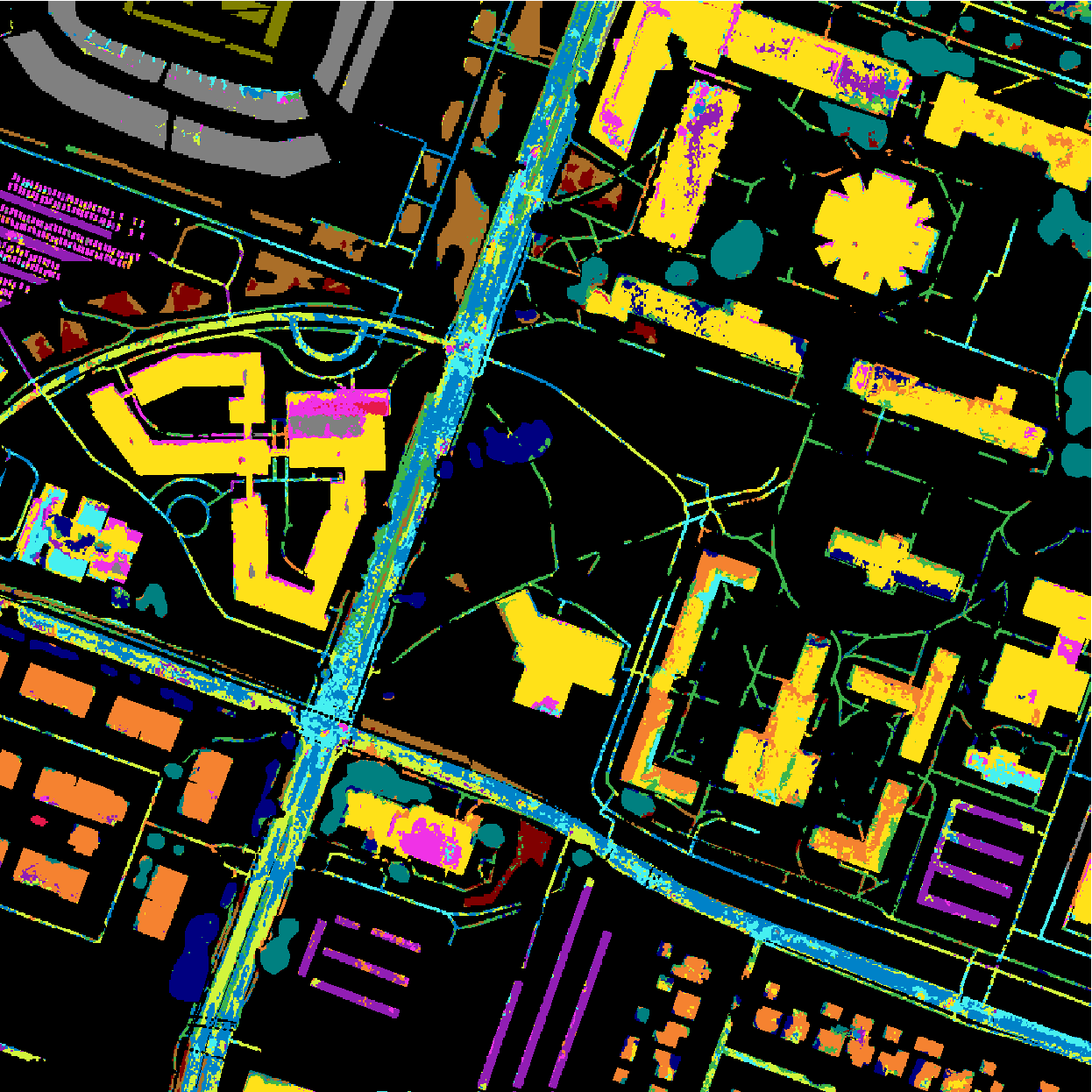}
&
\hspace{-10pt}
\includegraphics[width=0.32\linewidth]{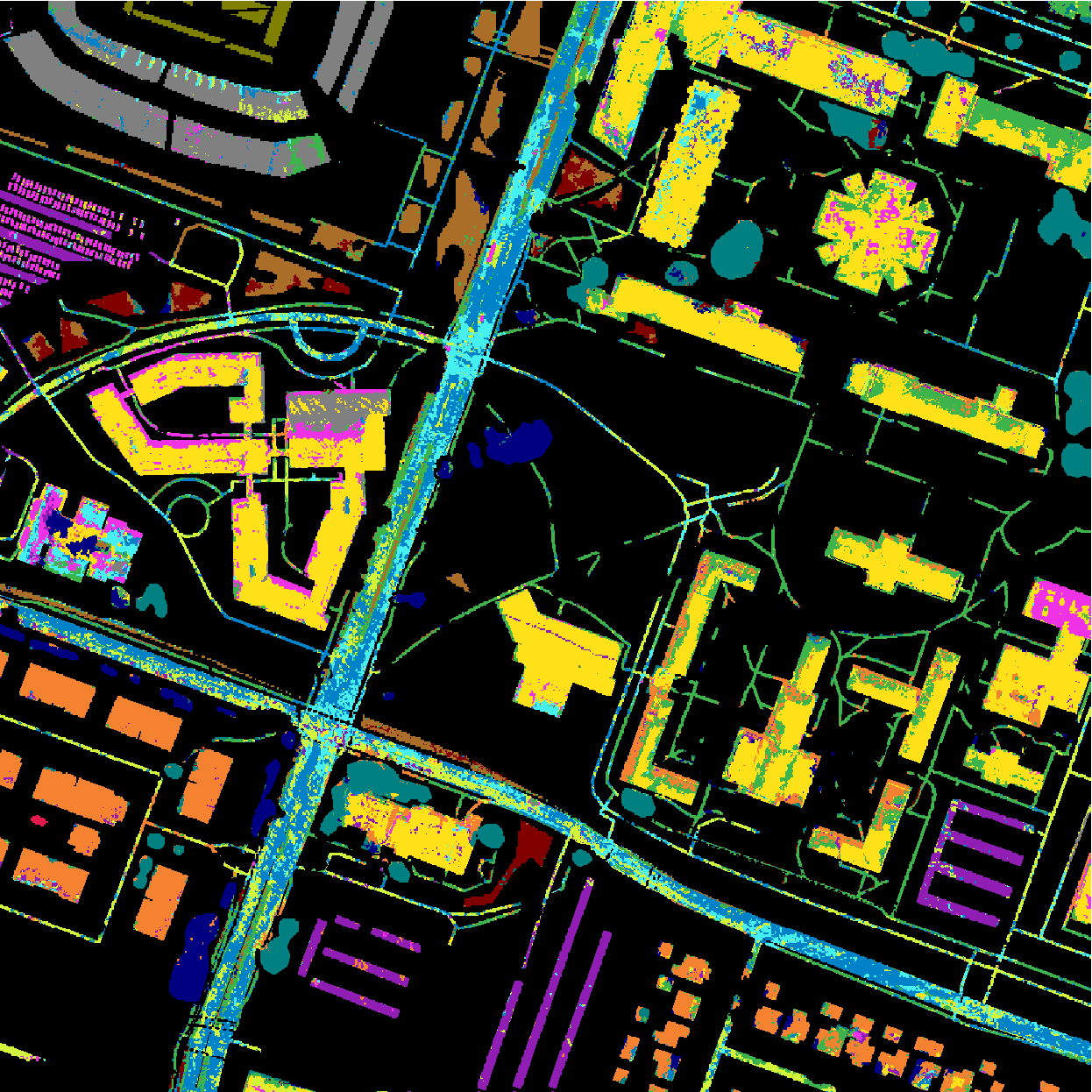}
\\
\hspace{-10pt} \scriptsize (d) RGB. OA:$45.73\%$
&
\hspace{-10pt} \scriptsize (e) HS. OA:$69.75\%$
&
\hspace{-10pt} \scriptsize (f) PCA. OA:$67.53\%$
\\
\hspace{-10pt}
\includegraphics[width=0.32\linewidth]{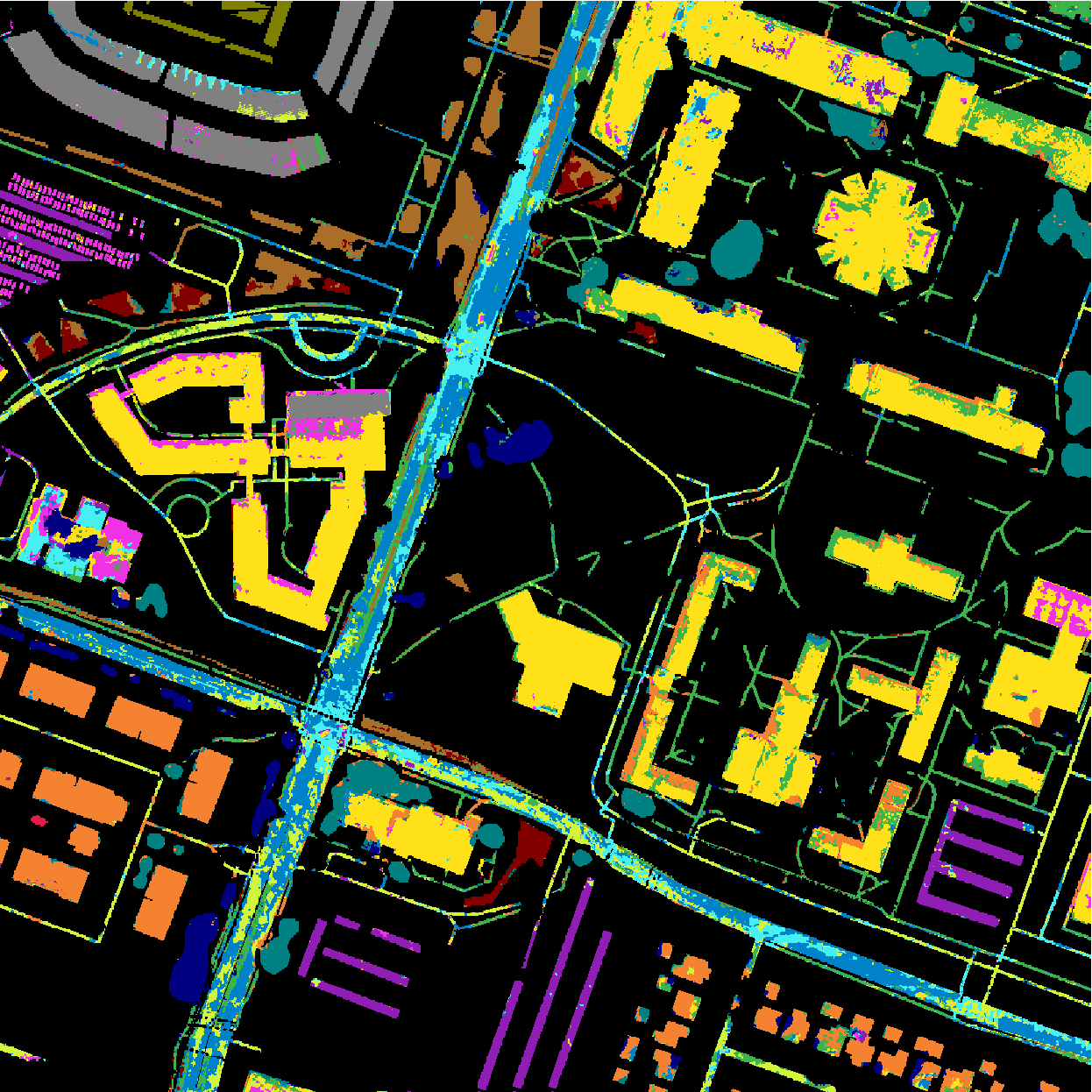}
&
\hspace{-10pt}
\includegraphics[width=0.32\linewidth]{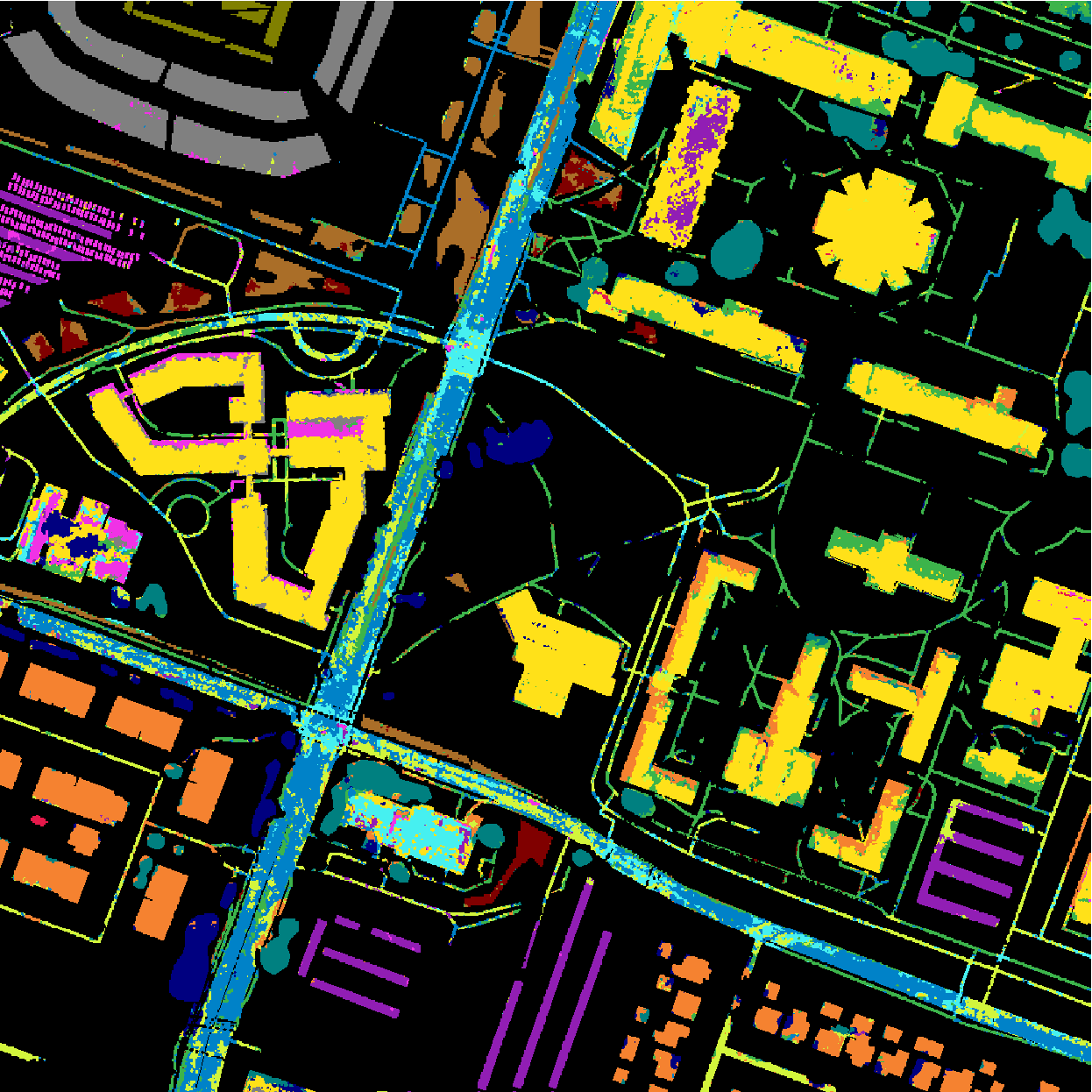}
&
\hspace{-10pt}
\includegraphics[width=0.32\linewidth]{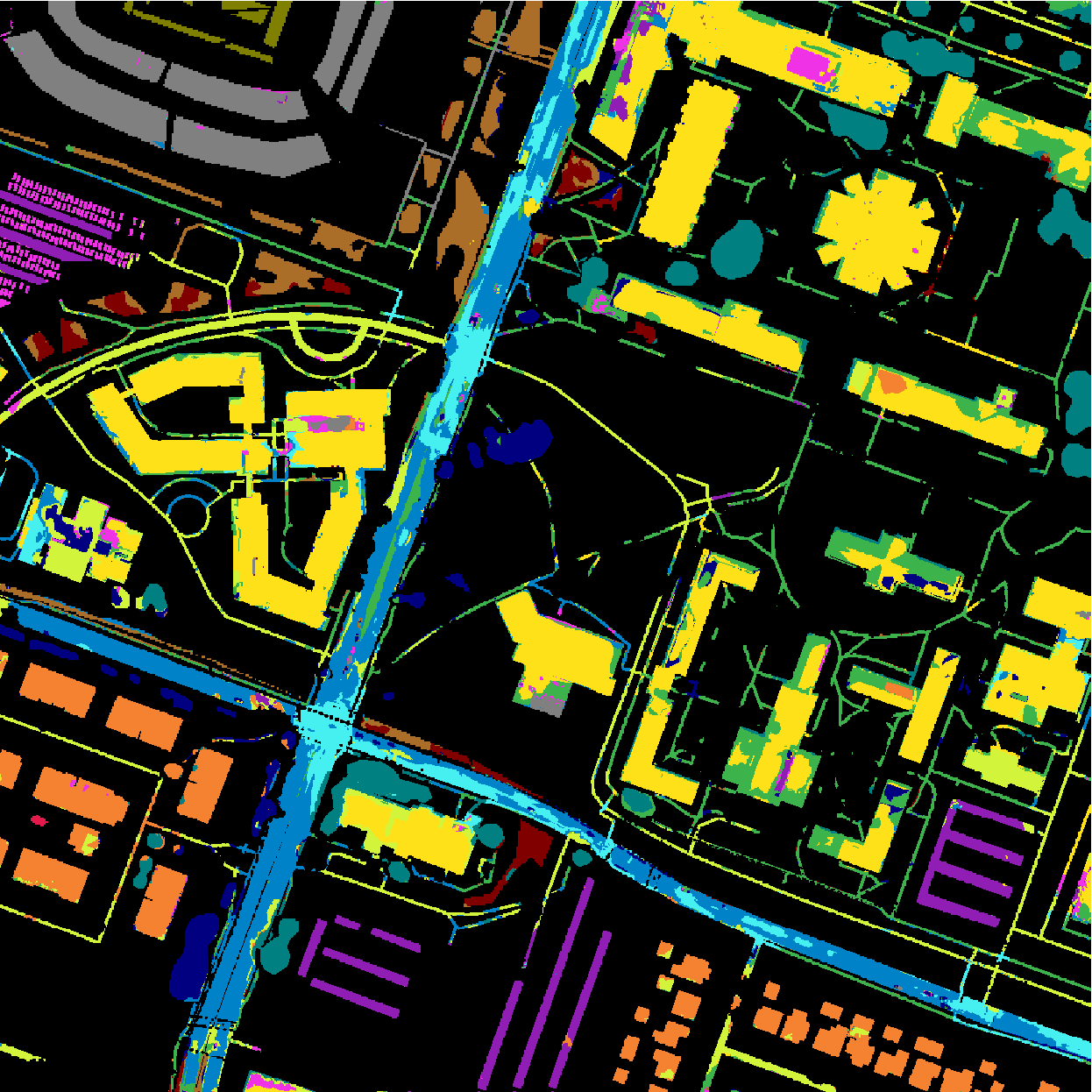}
\\
\hspace{-10pt} \scriptsize (g) SSLRA. OA:$72.85\%$
&
\hspace{-10pt} \scriptsize (h) SubFus. OA:$74.42\%$
&
\hspace{-10pt} \scriptsize (i) Proposed. OA:$75.26\%$
\end{tabular}\vspace{-15pt}
\end{center}
    \caption{\footnotesize Houston data set. (a) RGB image, (b) the RGB composite of the HS image, and (c) the ground truth map. (d)-(i) Labeling maps obtained by various the methods with their respective OA.}\vspace{-5pt}
    \label{fig:houston}
\end{figure}

\textbf{\textit{Houston data set}:} the proposed feature fusion technique was tested on a real data set. We used a cropped version of the Houston data set, also referred to as GRSS$\_$DFC$\_$2018 which was distributed for the 2018 GRSS Data Fusion Contest \cite{xu2019advanced}. The HS image was captured by an ITRES CASI 1500 system and it comprises $601 \times 596$ pixels and $48$ spectral bands in the wavelength range from 0.38 to 1.05 $\mu$m. Fig. \ref{fig:houston}(a) displays the RGB composite of the HS image. In addition, the RGB image was acquired by a DiMAC ULTRALIGHT sensor with a spatial resolution of 5 cm per pixel. In particular, we use a resized version of the RGB image with size 1202$\times$1192 that is illustrated in Fig. \ref{fig:houston}(b). Furthermore, the ground truth map with fifteen different classes is shown in Fig. \ref{fig:houston}(c).




We compare the performance of the feature fusion proposed method with respect to other classification approaches. In this sense, we first obtain the labeling maps from single sensor data. Figs \ref{fig:houston}(d) and \ref{fig:houston}(e) display the classification maps obtained from the RGB data and an interpolated version of the HS image, respectively. In order to consider a set of features obtained by fusing information from multi-sensor data, the PCA obtained from stacking the RGB image and an interpolated version of the HS image is determined. Then, a set of $N_e = 16$ PCA bands is selected to evaluate the classification performance. Fig. \ref{fig:houston}(e) shows the labeling map yielded by the PCA. We also apply the SSLRA method \cite{rasti2019hyperspectral} to the stacked data set whose classification map is illustrated in Fig. \ref{fig:houston}(f). Furthermore, Fig. \ref{fig:houston}(g) shows the labeling map yielded by the SubFus method \cite{rasti2020remote} from HS and RGB images. Finally, the labeling map obtained by the proposed feature fusion method is illustrated in Fig. \ref{fig:houston}(i). Parameter setting of the proposed feature fusion approach is fixed to $N_e = 16$ and $\lambda_{\mathrm{TV}} = 0.05$.


\setlength{\tabcolsep}{2pt}
\begin{table}
    \scriptsize
    \begin{center}
    \begin{tabular}{|c|l|c c||c c c c|}
    \hline
    \hline
     \multirow{2}{*}{Color} & \multirow{2}{*}{Classes} & \multicolumn{2}{c||}{$\#$ Samples} & PCA & SSLRA & SubFus & Proposed  \\
     & & Train & Test &  & \cite{rasti2019hyperspectral} & \cite{rasti2020remote} & \\
    \hline
    \hline
    \cellcolor[HTML]{800000} & Healthy grass & 75 & 5729 & 90.08 & 90.89 & 91.35 & \textbf{91.39} \\
    \cellcolor[HTML]{9A6324} & Stressed grass & 75 & 20055 & 85.06 & \textbf{86.64} & 85.00 & 85.91 \\
    \cellcolor[HTML]{808000} & Synthetic grass & 75 & 2639 & 99.51 & 99.57 & 99.51 & \textbf{100.00} \\
    \cellcolor[HTML]{469990} & Evergreen trees & 75 & 25387 & 94.45 & \textbf{94.78} & 93.80 & 93.90 \\
    \cellcolor[HTML]{000075} & Deciduous trees & 75 & 11652 & 93.76 & \textbf{94.62} & 94.27 & 94.33 \\
    \cellcolor[HTML]{E6194B} & Water & 75 & 55 & 98.18 & 99.55 & \textbf{99.64} & 99.27 \\
    \cellcolor[HTML]{F58231} & Residential & 75 & 31853& 83.74 & 87.49 & 91.59 & \textbf{96.42} \\
    \cellcolor[HTML]{FFE119} & Commercial & 75 & 191549 & 65.62 & 72.29 & \textbf{72.59} & 71.31 \\
    \cellcolor[HTML]{BFEF45} & Road & 75 & 27971 & 47.56 & 52.87 & 54.67 & \textbf{67.45} \\
    \cellcolor[HTML]{3CB44B} & Sidewalk & 75 & 49354 & 52.06 & 57.79 & 54.39 & \textbf{59.51} \\
    \cellcolor[HTML]{42D4F4} & Crosswalk & 75 & 1754 & 69.82 & 76.60 & 78.02 & \textbf{85.16} \\
    \cellcolor[HTML]{4363D8} & Major thoroughfares & 75 & 71007 & 50.95 & 55.32 & 58.90 & \textbf{65.32} \\
    \cellcolor[HTML]{911EB4} & Paved Parking & 75 & 16374 & 94.58 & 95.37 & 94.91 & \textbf{96.88} \\
    \cellcolor[HTML]{F032E6} & Cars & 75 & 4392 & 80.20 & 86.52 & 90.20 & \textbf{94.84} \\
    \cellcolor[HTML]{A9A9A9} & Seats & 75 & 26959 & 88.80 & 92.41 & \textbf{98.33} & 98.12 \\
    \hline
    \hline
    \multicolumn{4}{|c|}{Overall accuracy ($\%$)} & 68.11 & 72.93 & 73.82 & \textbf{76.00} \\
    \multicolumn{4}{|c|}{} & $\pm$1.52 & $\pm$1.41 & $\pm$1.64 & \textbf{$\pm$1.24}  \\
    \multicolumn{4}{|c|}{Average accuracy ($\%$)} & 79.63 & 82.85 & 83.81 & \textbf{86.65} \\
    \multicolumn{4}{|c|}{} & $\pm$0.50 & $\pm$0.51 & $\pm$0.35 & \textbf{$\pm$0.46} \\
    \multicolumn{4}{|c|}{Kappa Statistic} & 0.624 & 0.678 & 0.688 & \textbf{0.713} \\
    \multicolumn{4}{|c|}{} & 0.016 & 0.015 & 0.017 & \textbf{$\pm$0.014} \\
    \hline
    \hline
    \end{tabular}
    \end{center}
    \vspace{-10pt}
    \caption{\footnotesize Labeling accuracies yielded by the different feature extraction methods. \vspace{-15pt}}
    \label{tab:classifiers2}
\end{table}

To quantitatively evaluate the performance of the proposed method, Table \ref{tab:classifiers2} shows the classification accuracy obtained by the various feature fusion approaches. Specifically, every accuracy value is obtained by averaging 20 realizations of the respective experiment and at each trial, a different set of training samples is randomly selected. Furthermore, the overall accuracy, the average accuracy, and the Kappa statistic are shown in the last three rows of Table \ref{tab:classifiers2} for the various feature fusion techniques. Notice that the best accuracy values are shown in bold font. As can be observed in this table, the proposed method exhibit a competitive performance compared to the other feature fusion methods.

\vspace{-2mm}
\section{Conclusions}
\label{sec:conclusions}
\vspace{-2mm}

In this paper, a subspace-based feature fusion algorithm from HS and MS images was developed for land cover classification. In essence, the proposed approach extracts a set of high-resolution features that includes both the high-spatial-resolution information exhibited by the MS images and the rich spectral information embedded in HS images. To this end, the feature fusion model assumes that both the morphological profiles of the MS image and the HS image can be described as a high-spatial-resolution feature matrix lying in different subspaces. In this regard, a combination of alternating optimization (AO) and an ADMM algorithm was developed to solve the feature fusion problem. Note that this algorithm considers the degradation operators that describe HS and MS images to enhance the scene's spatial structure and thus, to improve the classification performance. In the experiments, we tested the labeling performance of the proposed approach for various parameter settings. Furthermore, it is also shown that the proposed feature fusion algorithm exhibits a competitive performance compared to other feature fusion methods.

\footnotesize
\bibliographystyle{IEEEbib}
\bibliography{strings,refs}

\end{document}